\documentclass[prl,aps,superscriptaddress,amssym,showpacs,nofootinbib,floatfix,twocolumn]{revtex4} 
\usepackage{epsfig,natbib}
\newcommand{\be}{\begin{equation}}
\newcommand{\ee}{\end{equation}}
\newcommand{\bea}{\begin{eqnarray}}
\newcommand{\eea}{\end{eqnarray}}
\newcommand{\ket}{\rangle}
\newcommand{\bra}{\langle}
\begin{document}

\title{Exact edge singularities and dynamical correlations in spin-1/2 chains}
\author{Rodrigo G. Pereira}
\affiliation{Department of Physics and Astronomy, University of British
  Columbia, Vancouver, BC, Canada V6T 1Z1}
\author{Steven R. White}
\affiliation{Department of Physics and Astronomy, University of California,
 Irvine CA 92697, USA}
\author{Ian Affleck}
\affiliation{Department of Physics and Astronomy, University of British
  Columbia, Vancouver, BC, Canada V6T 1Z1}
\date{\today}
\begin{abstract}

Exact formulas for the singularities of the dynamical structure factor, $S^{zz}(q,\omega)$, of the $S=1/2$ xxz spin chain at all $q$ and any anisotropy and magnetic field in the critical regime are derived, expressing the exponents in terms of the phase shifts which are known exactly from the Bethe ansatz solution. We also study the long time asymptotics of the self-correlation function $\bra0|S_j^z(t)S_j^z(0)|0\ket$. Utilizing these results to supplement very accurate time-dependent Density Matrix Renormalization Group (DMRG) for short to moderate times, we calculate $S^{zz}(q,\omega)$ to very high precision. 
\end{abstract}
\maketitle
The ``xxz'' $S=1/2$ spin chain, with Hamiltonian
\be H=J\sum_{j=1}^L[S^x_jS^x_{j+1}+S^y_jS^y_{j+1}+\Delta S^z_jS^z_{j+1}-hS_j^z],\label{xxzmodel}\ee
is one of the most studied models of strongly correlated systems. It is equivalent  by a Jordan-Wigner transformation to a model of interacting spinless fermions, with the corresponding Fermi momentum $k_F=\pi(1/2+\bra 0| S_j^z|0\ket)$ \cite{giamarchi}. The model with $\Delta=1$ describes Heisenberg antiferromagnets. The regime $0<\Delta<1$ is also of experimental interest; for example, the model with $\Delta=1/2$ can be realized in $S=1/2$ spin ladders near the critical field \cite{Watson}. In optical lattices, it should be even possible to tune the anisotropy $\Delta$ and explore the entire critical regime \cite{demler}.

While some aspects of the model have 
been solved for exactly by Bethe ansatz \cite{KorepinBOOK}, it has been very difficult 
to obtain correlation functions that way. Field theory (FT) methods 
give the low energy behavior at wave-vectors near $0$ and $2k_F$ \cite{giamarchi}.  From the experimental viewpoint \cite{Stone_neutron}, a relevant quantity  is the dynamical structure factor\be
S^{zz}(q,\omega)=\sum_{j=1}^Le^{-iqj}\int_{-\infty}^{+\infty}dt \,e^{i\omega t}\bra 0|S^z_j(t)S^z_0(0)|0\ket.
\ee  This is the Fourier transform of the density correlation function in the fermionic model. For $\Delta=1$ and $h=0$, the exact two-spinon contribution to $S^{zz}(q,\omega)$ was obtained from the Bethe ansatz \cite{Karbach}, partially agreeing with the M\"uller conjecture \cite{Muller}. More recently a number of new methods have emerged 
which now make this problem much more accessible.  These include 
time-dependent DMRG \cite{dmrg,WhiteFeiguin,FeiguinWhite}, calculation of form factors from Bethe ansatz \cite{Maillet,CauxMaillet} and new field theory approaches which go beyond the Luttinger model \cite{PustilnikPRL96, pereiraPRL}. The results point to a very nontrivial line shape at zero temperature for $S^{zz}(q,\omega)$ of the xxz model \cite{pereiraPRL} and of one-dimensional models in general \cite{PustilnikPRL96}. In the weak coupling limit $\Delta\ll 1$ and for small $q$, the singularities at the thresholds of the two-particle continuum have been explained by analogy with the x-ray edge singularity in metals \cite{PustilnikPRL96}. 

In this Letter we combine the methods of Ref. \cite{PustilnikPRL96} with the Bethe ansatz to investigate the singularity exponents of $S^{zz}(q,\omega)$ for the xxz model for finite interaction strength $\Delta$ and general momentum $q$. In addition, we determine the exponents of the long-time asymptotics of the spin self-correlation function, which  is \emph{not} dominated by low energy excitations. We check our predictions  against high accuracy numerical results  calculated by DMRG.

In the non-interacting, $\Delta=0$ case, only excited states with a single particle-hole pair contribute to $S^{zz}(q,\omega)$. All the spectral weight is confined between the lower and upper thresholds $\omega_{L,U}(q)$ of the two-particle continuum. The choices of momenta corresponding to the thresholds depend on both $k_F$ and $q$. For zero field, $k_F=\pi/2$,  $\omega_{L}(q)$ for any $q>0$ is defined by the excitation with a hole at $k_1=\pi/2-q$ and a particle right at the Fermi surface (or a hole at the Fermi surface and a particle at $k_2=\pi/2+q$), while $\omega_{U}$ is defined by the symmetric excitation with a hole at $k_1=\pi/2-q/2$ and a particle at $k_2=\pi/2+q/2$. For finite field and $q<|2k_F-\pi|$, $\omega_{L,U}(q)$ are defined by excitations with either a hole at $k_F$ and a particle at $k_F+q$ or a hole at $k_F-q$ and a particle at $k_F$. For $h\neq0$ and $q>|2k_F-\pi|$, there is even a third ``threshold" between $\omega_L$ and $\omega_U$ where $S^{zz}(q,\omega)$ has a step discontinuity (see \cite{Muller}).

For $\Delta\neq 0$,  $S^{zz}(q,\omega)$ exhibits a tail associated with multiple particle-hole excitations  \cite{pereiraPRL}. However, the thresholds of the two-particle continuum are expected to remain as special points at which power-law singularities develop \cite{PustilnikPRL96}. In order to describe the interaction of the high energy particle and/or hole with the Fermi surface modes, we integrate out all Fourier modes of the fermion field $\psi(x)$ except those near $\pm k_F$ and near the momentum of the hole, $k_1$, or particle, $k_2$, writing
\be \psi\left(x\right)\sim e^{ik_Fx}\psi_{R}+
e^{-ik_F x}\psi_{L}+e^{ik_1x}d_{1}+e^{ik_2x}d_2.\label{Psi}\ee
Linearizing the dispersion relation about $\pm k_F$ we obtain 
relativistic fermion fields which we bosonize in the usual way \cite{giamarchi}. We also expand the dispersion of the $d_{1,2}$ particles around $k=k_{1,2}$ up to quadratic terms. This yields the effective Hamiltonian density \bea
\mathcal{H} & = & \sum_{\alpha=1,2}d_{\alpha}^{\dagger}\left(\varepsilon_{\alpha}-iu_{\alpha}\partial_{x}-\frac{\partial_x^2}{2m_{\alpha}}\right)d^{\phantom{\dagger}}_{\alpha}\nonumber\\& &+\frac{v}{2}\left[\left(\partial_{x}\varphi_{L}\right)^{2}+\left(\partial_{x}\varphi_{R}\right)^{2}\right]+V_{12}d_1^{\dagger}d^{\phantom{\dagger}}_1d_2^{\dagger}d^{\phantom{\dagger}}_2\nonumber\\
 &  & +\frac{1}{\sqrt{2\pi K}}\sum_{\alpha=1,2}\left(\kappa^{\alpha}_{R}\partial_{x}\varphi_{R}+\kappa^{\alpha}_{L}\partial_{x}\varphi_{L}\right)d_{\alpha}^{\dagger}d^{\phantom{\dagger}}_{\alpha}.\label{effectiveH}\eea
This Hamiltonian describes a Luttinger liquid coupled to one or two mobile impurities \cite{Balents,Tsukamoto}. In the derivation of Eq. (\ref{effectiveH}) from Eq. (\ref{xxzmodel}), we drop terms of the form $(d^\dagger_{\alpha}d^{\phantom{\dagger}}_{\alpha})^2$ because we only consider processes involving a single $d_1$ and/or a single $d_2$ particle. Here $\varphi_{R,L}$ are the right and left components of the rescaled bosonic field. The long wavelength fluctuation part of $S^z_j$ is given by $S_j^z\sim \sqrt{K/2\pi}\, (\partial_x\varphi_R+\partial_x\varphi_L)$. The spin velocity $v$ and Luttinger parameter $K$ are known exactly from the Bethe ansatz \cite{KorepinBOOK}. For zero field, $v=(\pi/2)\sqrt{1-\Delta^2}/\arccos\Delta$ and $K=[2-2\arccos(\Delta)/\pi)]^{-1}$ (we set $J=1$). To first order in $\Delta$, the coupling constants describing the scattering between the $d$ particles and the bosons are $\kappa^{\alpha}_{R,L}=2\Delta[1-\cos(k_{F}\mp k_{\alpha})]$. The direct $d_1$-$d_2$ interaction $V_{12}$ is also of order $\Delta$. The exact values of $\kappa_{R,L}$ play a crucial role in the singularities and will be determined below.

We may eliminate the interaction between the $d$ particles and the bosonic modes by a unitary transformation
\be
U=\exp\left\{ i\sum_{\alpha}\int \frac{dx}{\sqrt{2\pi K}}\,\left(\gamma^{\alpha}_{R}\varphi_{R}-\gamma^{\alpha}_{L}\varphi_{L}\right)d_\alpha^{\dagger}d^{\phantom{\dagger}}_\alpha\right\},\label{U}\ee
with parameters $\gamma^{\alpha}_{R,L}=\kappa^{\alpha}_{R,L}/(v\mp u_\alpha)$. In the resulting Hamiltonian $\tilde{H}=U^\dagger HU$, $\varphi_{R,L}$ are free up to irrelevant interaction terms \cite{Balents}.  As in the x-ray edge problem, $\gamma^\alpha_{R,L}$ may be related to  the phase shifts at the Fermi points due to the creation of the high energy $d_\alpha$ particle.

Fortunately, we have access to the high energy spectrum of the xxz model by means of the Bethe Ansatz. Following the formalism of Ref. \cite{Tsukamoto}, we  calculate the finite size spectrum from the Bethe ansatz equations with an impurity term corresponding to removing (adding) a particle with dressed momentum $k_1=k(\lambda_1)$ ($k_2=k(\lambda_2)$), where $\lambda_{1,2}$ are the corresponding rapidities. The term of $O(1)$ yields $\varepsilon_\alpha=\epsilon(k_\alpha)$, the dressed energy of the particle. For zero field, we have the explicit formula $\epsilon(k)=-v\cos k$. The excitation spectrum for a single impurity to $O(1/L)$ reads
\bea
\Delta E&=&\frac{2\pi v}{L}\left[\frac{1}{4K}\left(\Delta N-n^\alpha_{imp}\right)^{2}+K\left(D-d^\alpha_{imp}\right)^{2}\right.\nonumber\\& &\qquad\left.+ n_{+}+n_{-}\right],\label{eq:spectrumBA}\eea
with a conventional notation for $\Delta N$, $D$ and $n_{\pm}$ \cite{KorepinBOOK}. The phase shifts $n^\alpha_{imp}$ and $d^\alpha_{imp}$ are given by 
\bea
n^\alpha_{imp}&=&\int_{-B}^{+B}d\lambda\, \rho^\alpha_{imp}(\lambda),\\
d^\alpha_{imp}&=&\int_{-\infty}^{-B}d\lambda\, \frac{\rho^\alpha_{imp}(\lambda)}{2}-\int_{B}^{+\infty}d\lambda\, \frac{\rho^\alpha_{imp}(\lambda)}{2},
\eea
where $B$ is the Fermi boundary and $\rho^\alpha_{imp}(\lambda)$ is the solution to the integral equation 
\be
\rho^\alpha_{imp}(\lambda)-\int_{-B}^{+B}\frac{d\lambda^\prime}{2\pi}\,\rho^\alpha_{imp}(\lambda^\prime)\frac{d\Theta(\lambda-\lambda^\prime)}{d\lambda}=\frac{\Phi^\alpha(\lambda)}{2\pi},\label{rhoimp}
\ee
where $\Theta(\lambda)=i\log[\sinh(i\zeta+\lambda)/\sinh(i\zeta-\lambda)]$, with $\Delta=-\cos\zeta$, is the two-particle scattering phase \cite{KorepinBOOK}, and $\Phi^{1,2}(\lambda)=\mp d\Theta(\lambda-\lambda_{1,2})/d\lambda$. The  spectrum of Eq. (\ref{eq:spectrumBA}) describes a shifted $c=1$ conformal field theory (CFT). The scaling dimensions of the various operators can then be expressed in terms of $K$, $n^\alpha_{imp}$ and $d^\alpha_{imp}$. In the effective model (\ref{effectiveH}), the shift is introduced by the unitary transformation of Eq. (\ref{U}), which changes the boundary conditions of the bosonic fields. The equivalence of the two approaches allows us to identify\be
\gamma^\alpha_{R,L}/\pi=n^\alpha_{imp}\pm2K d^\alpha_{imp}.
\ee
The phase shifts can be determined analytically for zero magnetic field. In this case, $B\to\infty$ and we have $d^\alpha_{imp}=0$. Moreover, by integrating Eq. (\ref{rhoimp}) over $\lambda$ we find \be
n^{1,2}_{imp}=\mp\Theta(\lambda\to\infty)/[\pi-\Theta(\lambda\to\infty)]=\pm(1-K).\label{phaseshift_h=0}
\ee

Once the exact phase shifts are known, the exponent for the (lower or upper) threshold determined by a single high energy particle can be calculated straightforwardly. For example, for a lower threshold defined by a deep hole, $\omega_L(q)=-\epsilon(k_F-q)$, the correlation function $\bra d_1^{\dagger} \psi^{\phantom{\dagger}}_R(t,x)\psi_R^{\dagger}d_1^{\phantom{\dagger}}(0,0)\ket$ can be factorized into a free $d_1$ propagator and correlations of exponentials of $\varphi_{R,L}$. After Fourier transforming, we find that near the lower edge  $S^{zz}(q,\omega)\sim \left[\omega-\omega_L(q)\right]^{-\mu}$ with exponent \cite{CP}\be
\mu=1-(1-n^1_{imp})^2/2K-2K(1/2-d^1_{imp})^2.\label{generalmu}
\ee 
For $h\to 0$, we use Eq. (\ref{phaseshift_h=0}) and obtain \be
\mu=1-K, \qquad (h\to0)
\ee
independent of the momentum of the hole. This form for the lower edge exponent had been conjectured long ago by M\"uller et al. \cite{Muller}. It agrees (up to logarithmic corrections) with the exponent of the two-spinon contribution to $S^{zz}(q,\omega)$ for the Heisenberg point  ($K=1/2$) \cite{Karbach}. 

The general result of Eq. (\ref{generalmu}) is consistent with the weak coupling expression for $\mu$ \cite{PustilnikPRL96}. To first order in $\Delta$, Eq. (\ref{generalmu}) reduces to \be
\mu\approx \frac{\kappa^1_R}{\pi(v-u_1)}\approx \frac{2\Delta\left(1-\cos q\right)}{\pi\left[\sin k_{F}-\sin\left(k_{F}-q\right)\right]}.
\ee
For $k_{F}\neq\pi/2$, we expand for $q\ll k_F$ and get $
\mu\approx m\Delta q/\pi$, 
where $m=(\cos k_F)^{-1}$ (\emph{c.f.} \cite{PustilnikPRL96}). For $k_{F}=\pi/2$, we obtain $
\mu\approx 2\Delta/\pi$, which is $1-K$ to  $O(\Delta)$. Note the cancellation of the $q$ dependence of $\kappa_R^1$ and $v-u_1$ in the latter case. Momentum-independent exponents have also been derived for the Calogero-Sutherland model \cite{PustilnikCSmodel}.
 
We now consider a threshold defined by high-energy particle and hole at $k_{1,2}=\pi/2\mp q/2$. The relevant correlation function is the propagator of the transformed $d_2^{\dagger}d_1^{\phantom{\dagger}}$. For simplicity, here we focus on the zero field case, in which $\varepsilon_2=-\varepsilon_1=v\sin(q/2)$, $u_2=u_1$ and $-m_2=m_1=[v\sin(q/2)]^{-1}$. Particle-hole symmetry then  implies that  $\gamma^1_{R,L}=\gamma^2_{R,L}$ and $d_2^{\dagger}d_1^{\phantom{\dagger}}$ is invariant under the unitary transformation of Eq. (\ref{U}).  In the noninteracting case, there is a square root singularity at the upper threshold due to the divergence of the joint density of states: $S^{zz}(q,\omega)\propto \sqrt{m_1/[\omega_U(q)-\omega]}$ for $\omega\approx\omega_U(q)=2v\sin(q/2)$ \cite{Muller}. For $\Delta\neq 0$, we need to treat the direct  interaction $V_{12}$ between the particle and the hole, which is not modified by $U$. This problem is analogous to the effect of Wannier excitons on the optical absorption rate of semiconductors \cite{Mahan,Ogawa}.  This simple two-body problem can be solved exactly for a delta function interaction. The result is that the upper edge exponent changes discontinuously for $\Delta\neq 0$: the square root divergence turns into a universal (for any $q$ and $\Delta$) square root cusp, $S^{zz}(q,\omega)\propto \sqrt{\omega_U(q)-\omega}$. This behavior contradicts the M\"uller ansatz \cite{Muller}, but is consistent with the analytic two-spinon  result for $\Delta=1$ \cite{Karbach}. Unlike the original exciton problem, a bound state only appears for $V_{12}<0$ ($\Delta<0$) \cite{boundstate}, because the particle and hole have a negative effective mass. For $\Delta\neq 0$, the upper edge cusp should intersect a high-frequency tail dominated by four-spinon excitations as proposed in \cite{JS4spinon}. This picture must be modified for $h\neq0$, since then $\gamma^1_{R,L}\neq\gamma^2_{R,L}$ and one needs to include the bosonic exponentials. The upper edge singularity then becomes $\Delta$- and $q$-dependent. The general finite field case, including the middle singularity \cite{Muller} for $q>|2k_F-\pi|$, will be discussed elsewhere.

We can apply the Hamiltonian of Eq. (\ref{effectiveH}) to study the self-correlation function $G(t)\equiv\bra 0|S^z_j(t)S^z_j(0)|0\ket $. Even in the noninteracting case, the long time asymptotics is a \emph{high energy} property, since it is dominated by a saddle point contribution with a hole at the bottom and a particle at the top of the band \cite{Sirker}. In this case, $k_1=0$ and $k_2=\pi$ and $d^{1,2}_{imp}$ vanish by symmetry ($\gamma^\alpha_{R}=\gamma^\alpha_{L}$). Here we restrict to zero field, but the method can be easily generalized. For $h=0$ and $\Delta\geq0$, $G(t)$ takes the form
\be
G(t)\sim B_{1}\frac{e^{-iWt}}{t^{\eta}}+B_{2}\frac{e^{-i2Wt}}{t^{\eta_{2}}}+\frac{B_{3}}{t^{\sigma}}+{B_4\over t^2}, \label{Gexp}
\ee
where $W=-\epsilon(0)=v$. The last two terms are the standard low-energy contributions, with $\sigma=2K$. The amplitudes $B_3$ and $B_4$ are known \cite{Lukyanov}. The first term is the contribution from the hole at the bottom of the band and the particle at $k_F=\pi/2$, with exponent
\be  \eta =(1+K)/2+(1- n^1_{imp})^2/2K=K+1/2.
\label{alpheta}\ee
The term oscillating at $2W$ comes from a hole at $k=0$ and a 
particle at $k=\pi$.  For $\Delta=0$, we have $\eta_2=1$. The exponent $\eta_2$ is connected with the singularity at the upper threshold of $S^{zz}(q,\omega)$ by $G(t)\sim \int d\omega e^{i\omega t}\int dq\, S^{zz}(q,\omega)$ for $q\approx\pi$ and $\omega\approx\omega_U(\pi)= 2v$. Due to the discontinuity of the exponent at $\omega_U$, $\eta_2$ jumps from $\eta_2=1$ to $\eta_2=2$ for any nonzero $\Delta$. This  behavior should be observed for $t\gg 1/(m_1V_{12}^2)\sim 1/\Delta^2$. As a result, the asymptotics of $G(t)$ is governed by the exponent $\eta<3/2$ for $0<\Delta<1$. For $\Delta<0$, we must add to Eq. (\ref{Gexp}) the contribution from the bound state.

We can also study $S^{zz}(q,\omega)$ with time-dependent DMRG (tDMRG) \cite{dmrg, WhiteFeiguin}. The
tDMRG methods directly produce $S^{zz}(x,t)$ and its spatial Fourier transform  $S^{zz}(q,t)$
for short to moderate times.  This information nicely complements the asymptotic information
available analytically. The DMRG calculation begins with the standard finite system calculation
of the ground state $\phi(t=0)$ on a finite lattice of typical length $L=$ 200-400, 
where a few hundred states are kept
for a truncation error less than $10^{-10}$. One of the sites at the center of
the lattice is selected as the origin, and the  operator $S^z_0$
is applied to the ground state to obtain a state $\psi(t=0)$. 
Subsequently, the time evolution operator
for a time step $\tau$, $\exp(i (H-E_0) \tau)$ where $E_0$ is the ground state
energy, is applied via a fourth order Trotter decomposition \cite{FeiguinWhite} to 
evolve both $\phi(t)$ and $\psi(t)$.  At each DMRG step centered
on site $j$ we obtain
a data point for the Green's function $G(t,j)$ by evaluating
$\langle \phi(t) | S^z_j | \psi(t) \rangle$.  As the time evolution progresses, the truncation
error accumulates. The integrated truncation error provides a useful estimate of
the error, and so longer times require smaller truncation errors
at each step, attained by increasing the number of states kept $m$. The truncation
error grows with time for fixed $m$, and is largest near the center where the
spin operator was applied. We specify
the desired truncation error at each step and choose $m$ to achieve it, within a specified
range. Typically for later times we have $m \approx 1000$. Finite size effects are small
for times less than $(L/2)/v$.
We are able to obtain very accurate results for $G(t,j)$, with errors between $10^{-4}$ and
$10^{-5}$, for times up to $Jt \sim$ 30-60.

\begin{figure}[b]
\includegraphics*[width=0.9\hsize,scale=1.0]{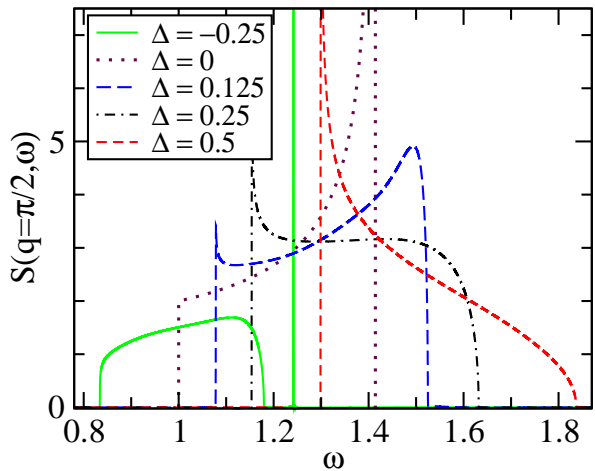}
\caption{(Color online).  DMRG results for $S^{zz}(q,\omega)$ versus $\omega$ for $q=\pi/2$, $h=0$ and several values of anisotropy $\Delta$. The line shapes for $\Delta>0$  show a divergent x-ray type lower edge and a universal square-root cusp at the upper edge. The curve for $\Delta<0$ shows a bound state above the upper edge. The width of the peak is very small for small $|\Delta|$. \label{fig1}
}
\end{figure}

For $Jt > 10-20$, we find the behavior of $S^{zz}(q,t)$ and $G(t)$ is well approximated by asymptotic
expressions, determined by the singular features of $S^{zz}(q,\omega)$ and $G(\omega)$. 
By utilizing the leading and subleading terms for each singularity, we have been able
to fit with a typical error in $S^{zz}(q,t)$ or $G(t)$ for $Jt \sim$ 20-30 between $10^{-4}$ and
$10^{-5}$.  We can fit with the  decay exponents determined analytically
or as free parameters to check the analytic expressions. Table \ref{tableII} shows the comparison between the exponents for $G(t)$ extracted independently from the DMRG data and the FT predictions.  In all cases the agreement
is very good. By smoothly transitioning from the tDMRG data to the fit as $t$ increases,
we obtain accurate results for all times. A straightforward time Fourier transform with a very long
time window yields very accurate high resolution spectra. Examples of line shapes obtained this way are shown in Fig. \ref{fig1}. We also did DMRG for the hole Green's function for the fermionic model corresponding to Eq. (\ref{xxzmodel}), obtaining good agreement with the predicted singularities from the x-ray edge picture.

\begin{table}[t]

\caption{Exponents for the spin self-correlation function $G(t)$ for $h=0$. The parameters $W$, $\eta$, $\eta_2$ and $\sigma$ were obtained numerically by fitting the DMRG data according to Eq. (\ref{Gexp}). These are compared with the corresponding FT predictions (with $v$ and $K$ taken from the Bethe ansatz). \label{tableII}}

\begin{tabular}{c|cc|cc|cc|cc}
\hline\hline 
$\Delta$&
$W$&
$v$&
$\eta$&
$\frac{1}{2}+K$&
$\sigma$&
$2K$&
$\eta_{2}$&

 \\
\hline 
0&
1&
1&
1.5&
1.5&
2&
2&
1&
1\\
0.125&
1.078&
1.078&
1.451&
1.426&
1.954&
1.852&
1.761&
2\\ 
0.25\phantom{0}&
1.153&
1.154&
1.366&
1.361&
1.811&
1.723&
2.034&
2\\ 
0.375&
1.226&
1.227&
1.313&
1.303&
1.694&
1.607&
2.000&
2\\
0.5\phantom{00}&
1.299&
1.299&
1.287&
1.25\phantom{0}&
1.491&
1.5\phantom{00}&
2.120&
2\\
0.75\phantom{0}&
1.439&
1.438&
1.102&
1.149&
1.324&
1.299&
2.226&
2\\
\hline\hline

\end{tabular}
\end{table}

We have not seen any exponential damping of the $\eta_2$ term in $G(t)$ for $\Delta>0$. This suggests that the singularity at the upper edge is not smoothed out in the integrable xxz model, even when the stability of the excitation is not guaranteed by kinematic constraints \cite{Khodas}. Integrability also protects the singularity at $\omega_U$ for finite field, as implied by the CFT form of the spectrum in Eq. (\ref{eq:spectrumBA}).

In conclusion, we presented a method to calculate the singularities of $S^{zz}(q,\omega)$ for the xxz model. The exponents for general anisotropy, magnetic field and momentum can be obtained by solving the Bethe ansatz equations which determine the exact phase shifts. For the particle-hole symmetric zero field case, we showed that the lower edge exponent  is $q$-independent and the (``exciton-like") upper edge has a universal square root singularity. The combination of analytic methods with the tDMRG overcomes the finite $t$ limitation on the resolution of the tDMRG and can be used to study dynamics of other one-dimensional systems (integrable or not).

We thank L. Balents, J.-S. Caux, V. Cheianov, L.I. Glazman, N. Kawakami, M. Pustilnik and J. Sirker for discussions. We acknowledge the support of the CNPq grant 200612/2004-2 (RGP), NSERC (RGP, IA), NSF under grant DMR-0605444 (SRW) and CIfAR (IA).


\begin{thebibliography}{99}
\bibitem{CP}After this paper was accepted we learned of 
the results of Cheianov and Pustilnik (private communication and 
arXiv:0710.3589). We have checked that their exponent is exactly the same 
as ours for all $q$, $\Delta$ and $h\neq 0$.
\end{thebibliography}
\end{document}